\documentclass[11pt]{article}
\usepackage{moriond,epsfig, graphicx}
\usepackage{multirow}
\pdfoutput=1
\def\MyPhoto{\pdfimage width 35mm {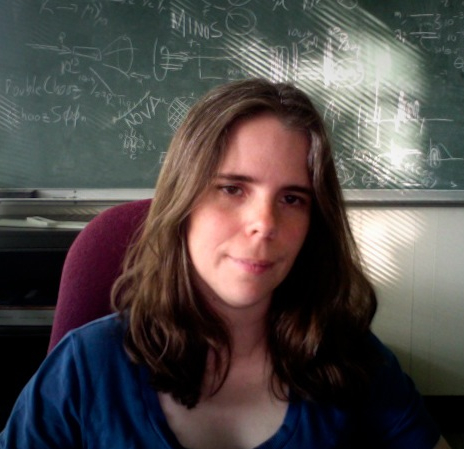}}

\def\Journal#1#2#3#4{{#1} {\bf #2}, #3 (#4)}

\def\PRL{\em Phys. Rev. Lett.}
\def\PRD{{\em Phys. Rev.} D}

\def\be{\begin{equation}}
\def\ee{\end{equation}}
\def\bea{\begin{eqnarray}}
\def\eea{\end{eqnarray}}

\newcommand{\annpred}{\ensuremath{27\pm 5 {\rm (stat.)} \pm 2 {\rm (syst.)}}}

\newcommand{\annobs}{\ensuremath{35}}

\newcommand{\delmsq}[1]{\ensuremath{\Delta m^2_{ #1 }}}
\newcommand{\dmsq}[1]{\delmsq{ #1 }}
\newcommand{\sinsq}[1]{\ensuremath{\sin^{2}\left(2\theta_{ #1 }\right)}}

\def\mdmatm{$\Delta m^2_{32}$}

\def\numunue{$\nu_\mu \rightarrow \nu_e$}

\def\numunutau{$\nu_\mu \rightarrow \nu_{\tau}$}

\newcommand{\numu}{\mbox{$\nu_{\mu}$}}                   
\newcommand{\nue}{\mbox{$\nu_{e}$}}                      
\newcommand{\nutau}{\mbox{$\nu_{\tau}$}}                 
\newcommand{\anue}{\mbox{$\overline{\nu}_{e}$}}          
\newcommand{\anumu}{\mbox{$\overline{\nu}_{\mu}$}}       

\newcommand{\piz}{\mbox{$\pi^{0}$}}                      

\begin{document}
\vspace*{4cm}
\title{Initial Results for Electron-Neutrino Appearance in MINOS}

\author{ M.C. Sanchez for the MINOS Collaboration }

\address{Argonne National Laboratory, High Energy Division, 9700 S. Cass Ave. \\
Argonne, Illinois 60439, USA}

\maketitle\abstracts{
MINOS is a long baseline neutrino oscillation experiment designed to make precision measurements of the neutrino mixing parameters associated with the atmospheric neutrino mass splitting. Using a neutrino beam from the Main Injector (NuMI) facility at Fermilab, it compares the neutrino energy spectrum for neutrino interactions observed in two large detectors located at Fermilab and in the Soudan mine in northern Minnesota at a distance of 735km. We describe initial results for electron neutrino appearance in MINOS after two years of data-taking. We observe \annobs{} events in the Far Detector with a background prediction of \annpred{} based on the measurement in the Near Detector.  The 1.5$\sigma$ excess of events can be interpreted in terms of \numu{}$\rightarrow$\nue{} oscillations. At 90\% CL we obtain an upper limit range of  $\sin^2(2\theta_{13})< 0.28-0.34$ for the normal neutrino mass hierarchy and $\sin^2(2\theta_{13})< 0.36-0.42$ for the inverted hierarchy depending upon $\delta_{CP}$.}

\section{Introduction}

Over the last decade, several experiments have provided compelling evidence for neutrino oscillations by detecting neutrinos originated in the atmosphere~\cite{ref:superk,ref:soudan} or in the sun~\cite{ref:superk,ref:sno}. These observations have been confirmed more recently by a subsequent generation of experiments based on artificial sources of neutrinos, such as accelerators~\cite{ref:k2knumu,ref:minos08} and reactors~\cite{ref:kamland}. The results support the description of three flavors of neutrinos changing identity as they travel and being related to three mass eigenstates as described by the Pontecorvo-Maki-Nakagawa-Sakata mixing matrix~\cite{ref:PMNS}.

The MINOS long baseline neutrino experiment at Fermilab has provided compelling evidence for \numu{} disappearance, establishing that neutrinos oscillate as they travel~\cite{ref:minos08}. MINOS has a baseline of 735~km and neutrinos peaking at 2-3~GeV, thus it is dominated by the atmospheric mass splitting of \mdmatm{}. MINOS, in fact, provides the most precise measurement for  this mass scale: $|$\mdmatm $| = 2.43 \pm 0.13 \times 10^{-3}$ eV$^2$. In this regime, the dominant oscillation channel is expected to be \numunutau, but it is possible that \numunue{} oscillations could occur. Observation of this oscillation channel would imply a non-zero value of the mixing parameter $\theta_{13}$, one of the missing pieces of the neutrino puzzle. A non-zero value of this parameter opens the door to the observation of CP violation of the neutrino sector. As of this writing, disappearance of the reactor \anue{} over a short baseline of few km has not been observed~\cite{ref:chooz}. This result implies that $\theta_{13}$ must be small, as measured by the Chooz experiment, $\sin^2(2\theta_{13})< 0.15$ at 90\% confidence level (CL) for the MINOS measured value of \mdmatm. Two other experiments have given limits, albeit with lower sensitivity~\cite{ref:k2k,ref:paloalto}. 

\section{The MINOS Experiment} 
In MINOS, neutrino interactions from the Fermilab Main Injector (NuMI) facility~\cite{ref:numi} are recorded at two detectors: Near   Detector (ND), 1~km from the NuMI target, and a Far Detector (FD) at 735~km from the same target. The detectors are tracking calorimeters composed of 2.54~cm thick steel absorber planes and 1.0~cm thick active scintillator planes forming detector layers that correspond to $\sim 1.4$ radiation lengths.  The scintillator planes are composed of 4.1~cm wide strips which corresponds to $\sim 1.1$ Molier\`e radii.  The data recorded in the ND establishes the properties of the beam before oscillations have occurred~\cite{ref:minos08,ref:minosnim}. Evidence for oscillations is observed as distortions of the beam spectrum or composition measured at the FD with respect to the ND. Thanks to the similarity between the two detectors, systematic errors arising from uncertainties in the neutrino interactions or in the neutrino flux largely cancel. 

The neutrino beam is produced by impinging 120~GeV protons from the Fermilab Main Injector upon a graphite target, producing pions and kaons, which are then focused by two magnetic horns~\cite{ref:numi}.  The horn current and position of the target relative to the horns can be configured to produce different neutrino energy spectra. Most of the physics data has been acquired in the low energy configuration with a peak at 2-3~GeV, but other configurations have been used to study the backgrounds and systematics. The beam is 98.7\% \numu{} and \anumu{} and 1.3\% intrinsic \nue+\nue. The latter originate from decays of muons produced in pion decays and from kaon decays. The muon component dominates below 8~GeV and it is well constrained by the \numu-CC flux from different beam configurations~\cite{ref:minos08,ref:minosprd}. Errors on the \nue{} content of the beam are 3.5\% at the ND and 5.7\% at the FD. 

The FD data for the analysis presented here was recorded between May 2005 and July 2007, corresponding to an exposure of $3.14\times 10^{20}$ protons on target (POT). The ND data are sampled at a 1/6 rate uniformly over the 2 year run period. The data are partitioned in two subsets, with the second data set acquired with a $\sim 1$~cm longitudinal offset relative to the target position of the first data set. This offset results in a $\sim 30$~MeV shift in the neutrino spectrum that has been incorporated in the Monte Carlo (MC) simulation appropriately weighted by exposure. The MINOS MC simulation of the beam line and detectors is based on {\tt GEANT3}~\cite{ref:geant}. The simulation of the hadron production in the target is done using {\tt FLUKA}~\cite{ref:minosnim,ref:fluka}. The {\tt NEUGEN} simulation is used to model the neutrino interactions and it has been tuned to available bubble chamber data from relatively higher invariant masses than the region of interest for MINOS. The fragmentation or hadronic shower model uses the KNO description for the low invariant mass and transitions gradually at higher masses to the JETSET model~\cite{ref:neugen}.

\section{Finding Electron Neutrinos in MINOS}

The flavor of the incoming neutrino can only be identified in charged current (CC) neutrino interactions, when the charged lepton partner of the incoming neutrino is produced in the final state. The 
MINOS detectors were designed to preferably detect long muon tracks resulting from muon-neutrino charged current interactions $\numu{}+{\rm N}\rightarrow \mu+X$ (\numu-CC). However, one can identify electron-neutrino charged current interactions $\nue{}+{\rm N}\rightarrow e+X$ (\nue-CC) by searching for electrons which deposit their energy in a narrow and relatively short  span in the MINOS calorimeters with a longitudinal distribution as described by the gamma function~\cite{ref:PDG}. Additional calorimeter activity can be produced by the hadronic showers resulting from the breakup of the recoil nucleus, $X$. There are other neutrino scattering processes that can produce similar topologies in the MINOS detectors, such as neutral current (NC) interactions $\nu +{\rm N}\rightarrow \nu+X$, as well as \numu-CC interactions where little energy is transferred to the out-going muon. Both of these background components are dominated by hadronic showers, typically possessing an electromagnetic element arising from \piz{} decays. Other less significant backgrounds, arise from beam \nue-CC interactions, \nutau-CC interactions from oscillations and cosmogenic sources. 

The data are first filtered to ensure data and beam quality. Fiducial volume cuts are imposed to assure well reconstructed events. In order to select a \nue-CC like event sample, we restrict the energy range between 1 and 8~GeV, which contains the \numunue{} oscillation probability maximum. The lower cut removes mainly NC events, while the higher cut removes beam \nue-CC events. Background from cosmogenic sources are suppressed to less than 0.5 event (at 90\% CL) by selecting events in time with the accelerator beam pulse and in the direction of the beam. Events are required to have a reconstructed shower and at least 5 contiguous planes with energy deposition greater than 0.5~MIP. Events with tracks longer than 25 planes or 12 track like planes are rejected. MC simulations indicate that these cuts improve signal to background ratio from 1:55 to 1:12 assuming a \sinsq{13} value at the CHOOZ limit. Two selection algorithms have been developed to enhance this ratio for higher background rejection, they are the ANN and LEM methods. The two selection algorithms rely on very different techniques, provide different signal to background ratios, and are sensitive to different systematic uncertainties. In the analysis reported here, the ANN selected sample is used to derive the final results but the LEM selection is presented as a cross check.

\subsection{The ANN selection method}
The final \nue{}-selected sample is obtained using a method based on an Artificial Neural Network (ANN) with 11 variables characterizing the longitudinal and transverse energy deposition profiles~\cite{ref:TJthesis}.  Some of these quantities are the energy fraction in windows of 2, 4 or 6 planes, the fraction of energy in a 3 strip wide road, the RMS of the transverse energy deposition, etc. The neural network is trained to separate the signal \nue{}-CC events from NC or \numu{}-CC background. The ANN generates a single numerical output with higher values indicating likely \nue{}-CC events. The acceptance cut value is chosen to be at 0.7 by maximizing the ratio of the accepted signal to the expected statistical and systematic uncertainty of the background. Assuming a signal at the CHOOZ limit, this method gives a 1:4 signal to background ratio, with a 41\% signal efficiency and a rejection efficiency at 92.3\% and 99.4\% for NC and \numu-CC respectively. 

\subsection{The LEM selection method}

A second selection method, the Library Matching Method (LEM), is used as a cross check. In this novel technique, each candidate is compared to a large library of simulated \nue-CC and NC events~\cite{ref:POthesis}. The 50 best matches are identified based on the relative probability that the hit pattern in a library and a candidate event come from the same neutrino interaction. Three variables are constructed: the fraction of these matches that are \nue-CC events, the mean hadronic $y$ of the best matches, and the mean fractional charge matched within those best matches. These quantities are then used as a function of energy to calculate a likelihood function that discriminates between signal and background. The acceptance cut value is chosen at 0.65. The LEM method gives better background rejection than the ANN, with a signal to background ratio of 1:3, but also displays increased sensitivity to certain systematic uncertainties. This method results in a  46\% signal efficiency and a rejection efficiency at 92.9\% and 99.3\% for NC and \numu-CC respectively.

\section{Estimating the Background for Electron-Neutrino Appearance}

Data in the ND are recorded before oscillations have occurred, thus events selected there by the ANN are background events. The data recorded in the ND are directly used to predict the number of background events expected in the FD. The background measured in the ND is comprised of three different components: NC events, \numu{}-CC events and beam \nue{} events.  Oscillations and beam line geometry considerations require that each component be treated separately in the prediction of the FD backgrounds. Figure~\ref{fig:nddata} (left) shows the energy spectrum of events selected with the ANN as \nue{}-CC like for data and MC in the ND. The data agrees with the simulation within the uncertainty which is dominated by the hadronic shower model systematic errors. 

\subsection{Background estimation using horn-off data}

\begin{figure}
\begin{center}

\[ \pdfimage height 2.5in {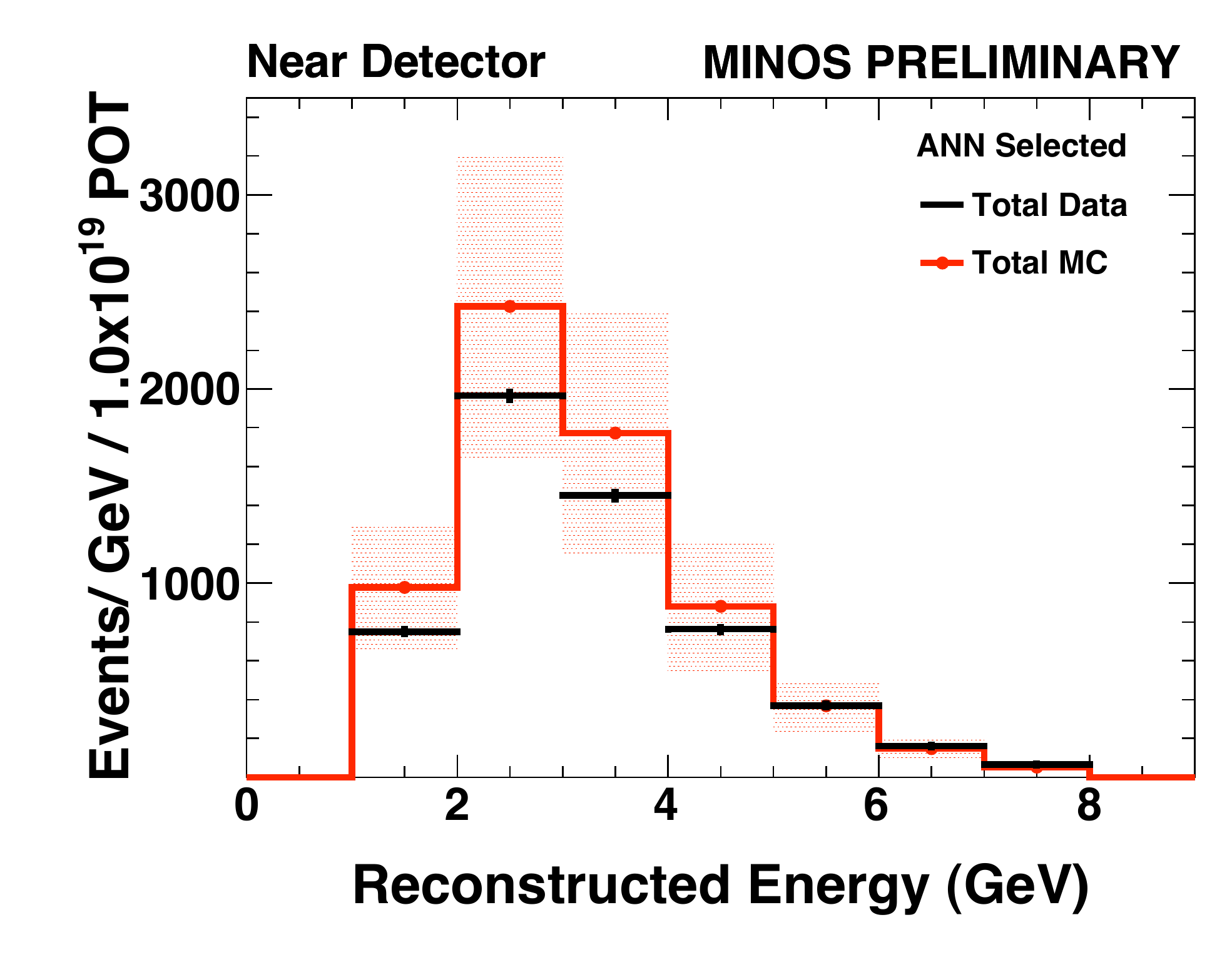} 
\pdfimage height 2.5in {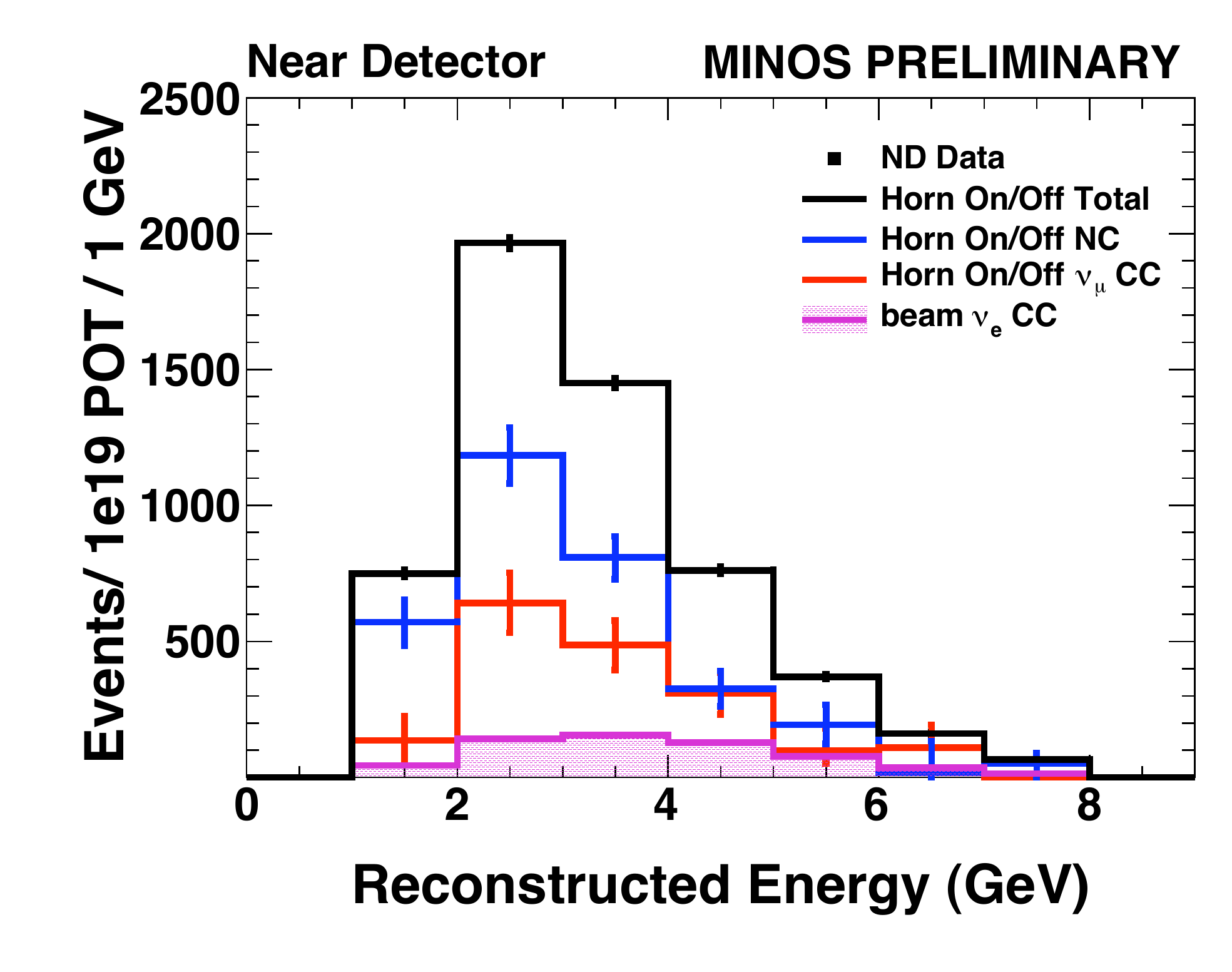}
\]
\caption{Reconstructed energy spectrum of \nue-selected ND data with total MC (left). The black points show data with statistical errors, the red histogram shows the MC with systematic errors as a shaded region. Reconstructed energy spectrum of \nue-selected ND data with predicted background components (right). The solid histogram corresponds to the total of the 3 background components present in the ND data and it has been tuned to agree with the data points. The selected background from NC is shown in blue and from \numu-CC in red, both as obtained by the horn-off method. The shaded histogram shows the beam \nue-CC component from the MC simulation. The errors in the data are statistical and are obscured by the markers; the errors on the components are systematic. 
\label{fig:nddata}}
\end{center}
\end{figure}

The background components are determined using a NC-enriched data sample recorded with the focusing horns turned off (horn-off). In this configuration the pions are not focused, the low energy peak of the neutrino energy distribution disappears, leaving an event sample dominated by higher energy NC events. These data used in conjunction with the standard beam configuration (horn-on) and the simulated horn-off to horn-on ratios allow us to extract the individual NC and \numu-CC components as a function of energy. The beam \nue-CC are used as an input for this method and are well constrained by the ND \numu-CC data. The ratios are shown to be well modeled before the application of the \nue-CC selection and uncertainties such as the hadronic shower modeling cancel to first order in these ratios.  Figure~\ref{fig:nddata} (right) shows the NC, \numu-CC, and intrinsic beam \nue-CC components of the background in the ND derived used this technique. Systematic errors on the components arise from uncertainties in the beam flux, cross section and selection efficiency and are derived from the data for the main background components, NC and \numu-CC. 

\subsection{Background estimation using muon-removed data}

A second technique allows us to study the ND measured background using an independent sample of showers derived from \numu-CC events selected with long tracks~\cite{ref:minos08}. These events are processed to remove the hits associated with the muon track~\cite{ref:JBthesis,ref:AHthesis}. Then the shower remnant is reconstructed to provide an independent sample of hadronic shower events. The procedure is done on data and MC and the \nue{} selection is applied to both. The ratio of these muon-removed data to muon-removed MC is used to correct the NC component of the background. The residual difference between data and MC is absorbed into the \numu-CC component by making the total MC agree with the data.  The predictions for the background components from this method agree well with those obtained from the horn-off method.

\begin{table} [t]
\caption{Breakdown of systematic uncertainties contributed by different sources to the total number of background events in the Far Detector.}
\begin{center}
\begin{tabular}{l l c}
\hline \hline 
\multirow{2}{*}{Uncertainty source} & & Uncertainty on\\
 & &  background events \\
\hline 
Far/Near ratio (systematic) & &  6.4\% \\
Horn-off (systematic) & & 2.7\% \\
Horn-off (statistical) & & 2.3\% \\

\hline
Total Systematic Uncertainty && 7.3\% \\
Expected Statistical Uncertainty & & 19\% \\
\hline \hline

\end{tabular}
\end{center}
\label{tab:systematics}
\end{table}

\section{Far Detector Signal and Background Prediction}

Once the ND energy spectrum is decomposed into its background components, each of these spectra is multiplied by the Far to Near ratio from the MC simulation for that component providing a prediction of the FD spectrum. The MC simulation takes into account differences in the spectrum of events at the ND and FD due to beam line geometry as well as differences in detector calibration and topological response. Oscillations are included when predicting the \numu-CC component. The smaller \nutau-CC and beam \nue-CC components are derived from the \numu-CC selected FD event energy predicted spectrum. For the ANN selection, we expect 26.6 background events, of which 18.2 are NC, 5.1 are \numu-CC, 2.2 are beam \nue-CC and 1.1 are \nutau-CC; using \dmsq{32}=$2.43\times10^{-3} {\rm eV^{2}}$, \sinsq{23}=1.0, and \sinsq{13}=0 as oscillation parameters.

It is possible to estimate the efficiency for selecting \nue-CC events by using the sample of muon-removed events and embedding a simulated electron of the same momentum as the muon. Test beam measurements indicate that the ANN selection efficiency on single electrons agrees with the simulation to within 2.6\% in the MINOS detectors~\cite{ref:JBthesis,ref:caldet}. Comparisons between the muon-removed (with added electron) data and MC simulation samples indicate that the selection efficiency of the signal events is well modeled. The \nue-selection algorithms focus on the electromagnetic core of the shower and seem not to be affected by the hadronic shower. The difference in selection efficiency between data and MC is -0.3\% and is used as a correction to the signal selection efficiency. 

The systematic errors were evaluated by generating modified MC samples and quantifying the change in the number of predicted background events in the FD~\cite{ref:JBthesis}. Table~\ref{tab:systematics} shows that the dominant uncertainty arises from Far/Near differences such as the relative energy scale calibration, the details of the modeling of the photomultiplier tubes and the relative normalization. Other uncertainties arising from neutrino interaction physics, shower hadronization, intranuclear re-scattering, and absolute energy scale errors largely cancel out in the extrapolation. The Far/Near systematic errors are added in quadrature along with the systematic and statistical error arising from the background decomposition in the ND. As shown, the error is dominated by the expected statistical uncertainty on the number of background events for this data set. 

\begin{figure} [t]
\begin{center}

\[ \pdfimage height 2.5in {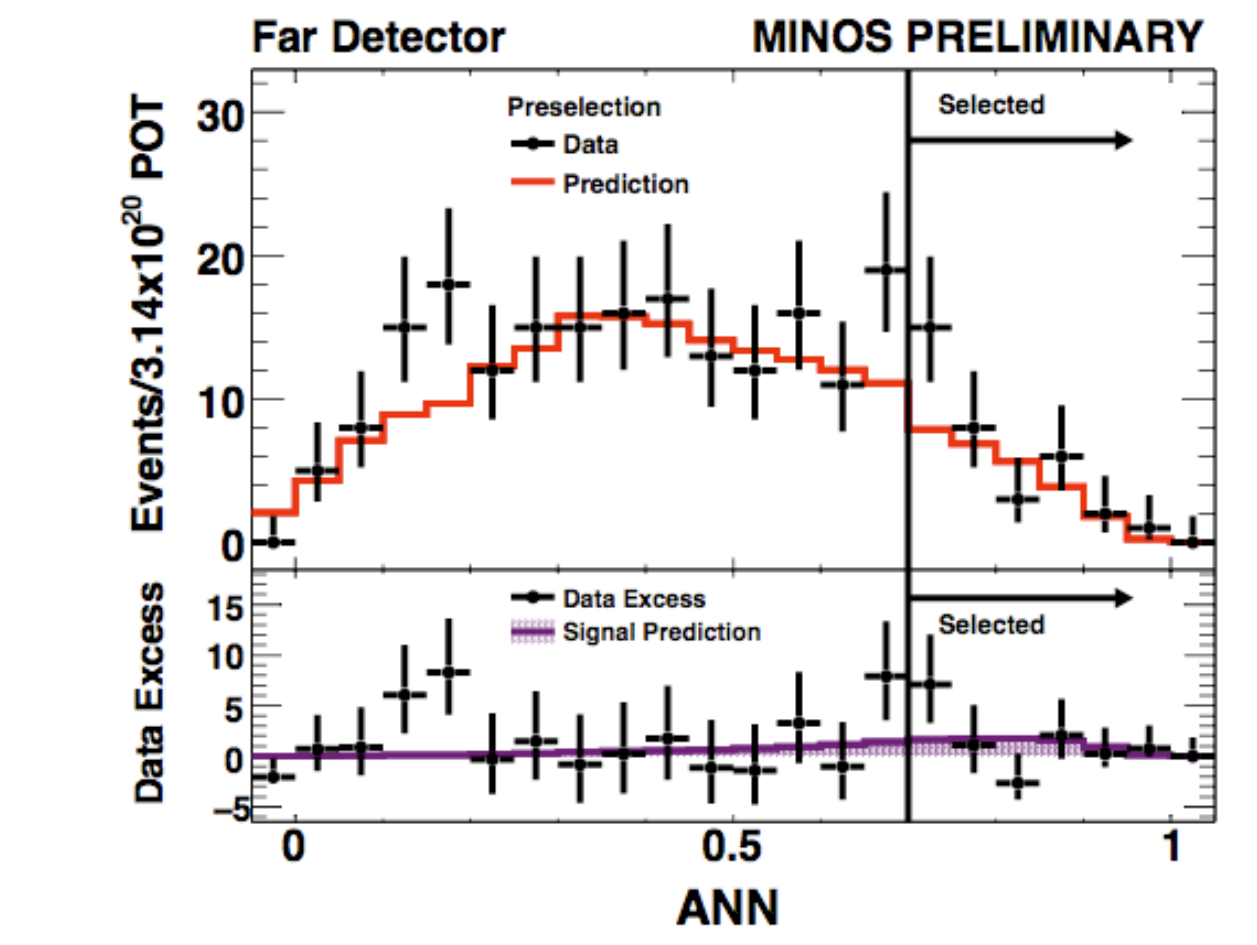} 
\pdfimage height 2.5in {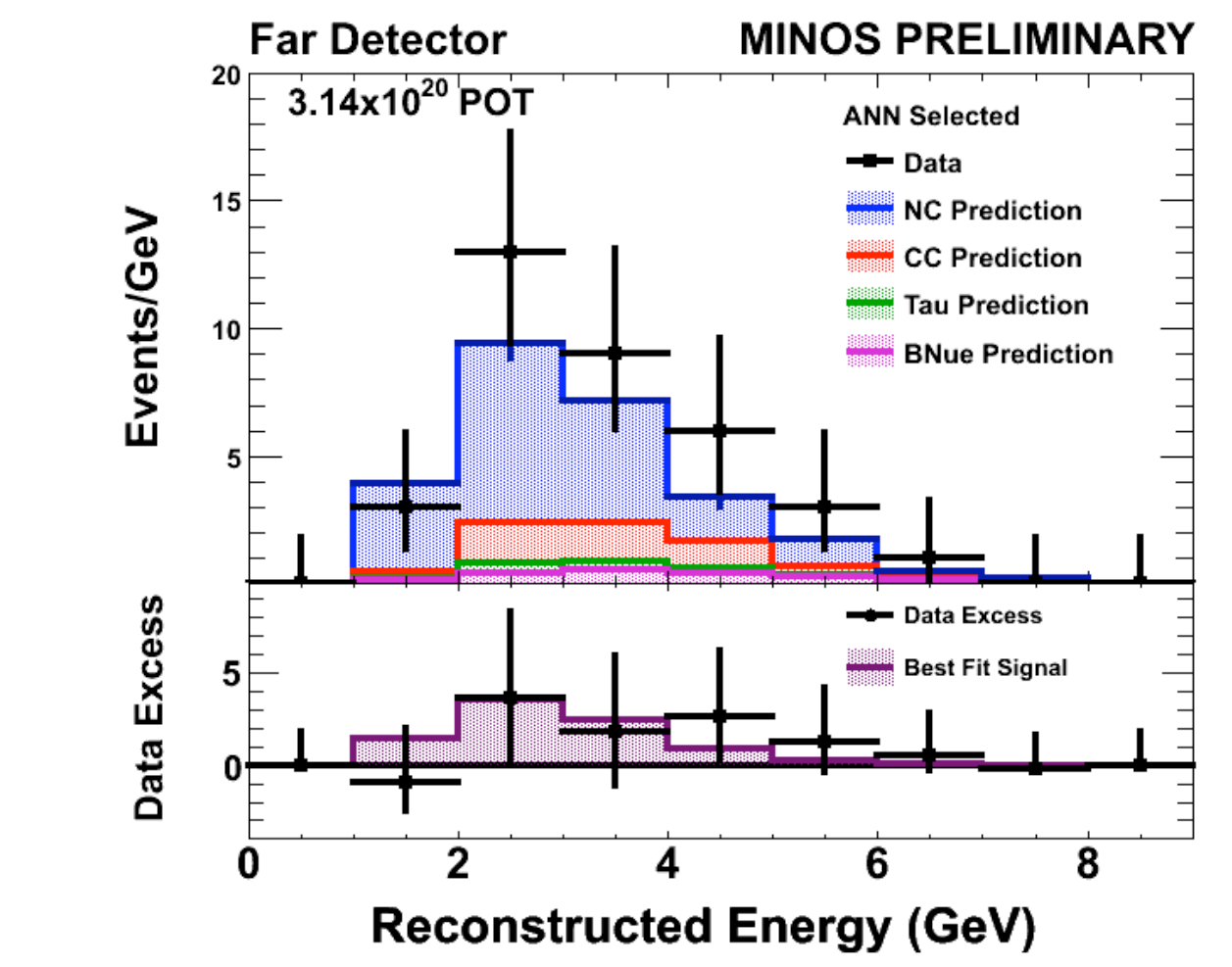}
\]
\caption{Distribution of the ANN selection variable for data and MC events in the FD (left). Black points show data with statistical error bars. The red histogram shows the background expectation. Below, the purple shaded histogram shows the \nue-CC prediction by the oscillation hypothesis compared to the data excess (black points). The reconstructed energy distribution of the \nue-CC selected events in the FD (right). Black points show the data with the statistical errors bars. The stacked histogram shows the total predicted background, subdivided in components: NC (blue), \numu-CC (red), \nutau-CC (green) and beam \nue-CC (magenta). Below, the purple shaded histogram shows \nue-CC prediction by the oscillation hypothesis compared to the data excess (black points). 
\label{fig:fddata}}
\end{center}
\end{figure}

\section{Far Detector Data}
We established the prediction of the background components and the signal as well as the systematic uncertainties before examining the data in the FD. In addition, we studied a signal free sample to validate the \nue{} selection methods and background estimation techniques we have developed. All FD data was studied with both \nue-CC selection methods. 

\subsection{Results using the ANN method}
The muon removal procedure was applied to the FD data and MC simulation. The number of events passing ANN selection was compared to the prediction based on the same sample from the ND. In this muon-removed sample, we observe 39 events above the selection cut, with an expectation of $29\pm 5$(stat.)$\pm2$(syst.). The excess of events is concentrated at very high values of the selection variable. Examination of the events and distributions give no evidence that the excess is not a fluctuation. The sample will be explored further with the larger data set currently being acquired. 

The standard FD data sample was first studied for events passing all but the  \nue-CC selection criteria. In the FD data, 146 events were observed below ANN=0.55, compared to a pure background expectation of $132\pm 12$(stat.)$\pm8$(syst.). In the signal region above ANN=0.7, we observe 35 events, with a background expectation of $27\pm 5$(stat.)$\pm2$(syst.). Figure~\ref{fig:fddata} (left) shows the ANN selection variable distribution. The observed energy spectrum for events within the signal region is shown in Figure~\ref{fig:fddata} (right). 

\subsection{Results using the LEM method}
The second selection method, LEM, was used as a cross check and it is not described in detail here. For the muon removed-sample using this selection method, we observe 25 events, with an expectation of $17\pm 4$(stat.)$\pm2$(syst.). For the standard FD data in the region below a cut of LEM=0.55, we observe 176 events, with an expectation of $157\pm 13$(stat.)$\pm3$(syst.). As in the ANN method, we observe a small excess ($< 2\sigma$) for both the muon-removed sample and the region below the selection cut. In the signal region, we observed 28 events compared to a background expectation of $22\pm 5$(stat.)$\pm3$(syst.). These results are also consistent with the ANN selection.

\section{Results for \numunue{} in MINOS}

\begin{figure}
\begin{center}

\[ \pdfimage height 2.8in {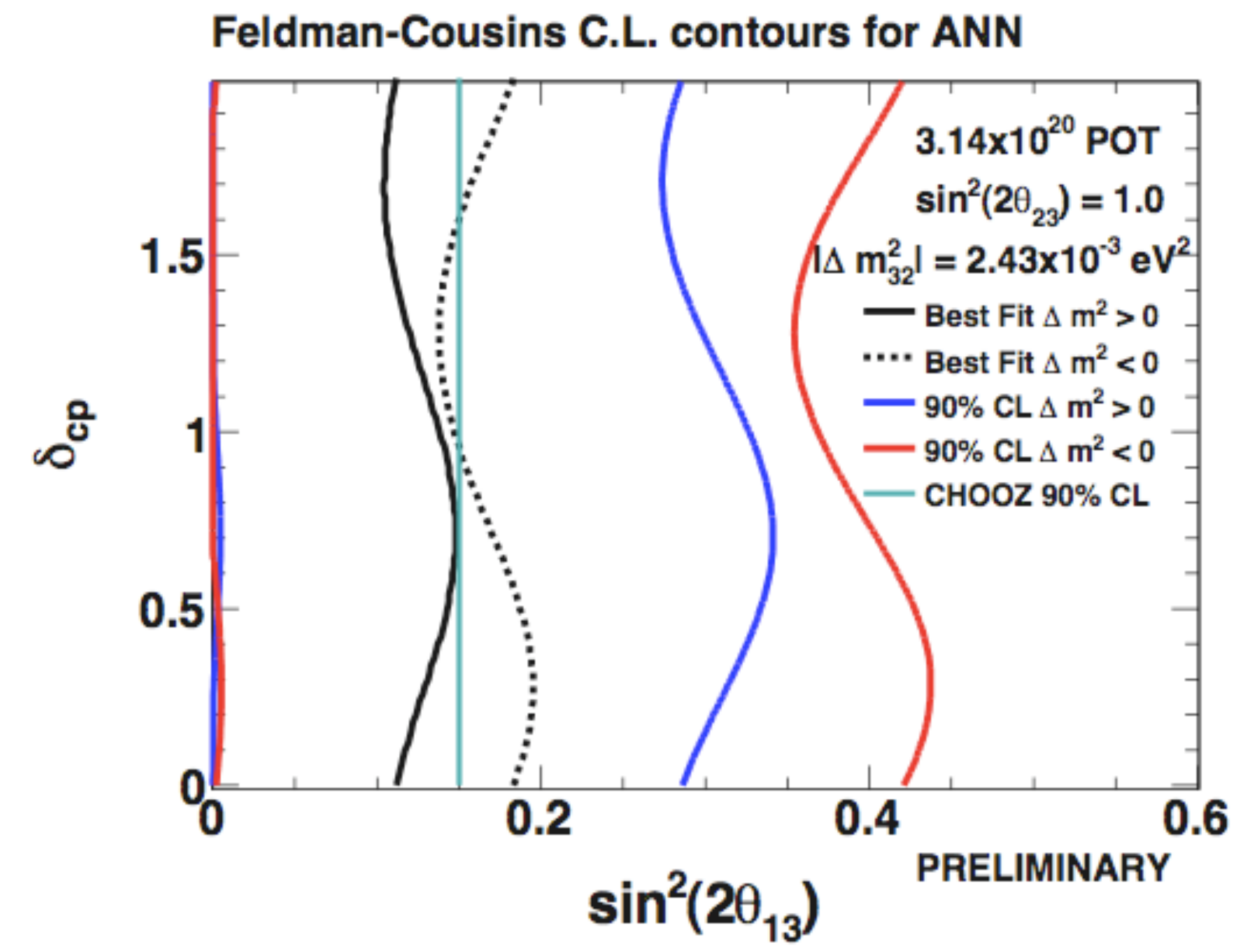} 
\]
\caption{The 90\% CL in the \sinsq{13}  and $\delta_{CP}$ plane using the ANN selection method. Black lines show the best fit to the data for both the normal hierarchy (solid) and inverted hierarchy (dotted). Blue (red) lines show the 90\% CL boundaries for the normal (inverted) hierarchies.
\label{fig:contour}}
\end{center}
\end{figure}

Figure~\ref{fig:contour} shows the 90\% confidence intervals in the \sinsq{13}  and the CP phase, $\delta_{CP}$, plane obtained using the observed total number of events selected by the ANN method. The oscillation probability is computed using a full 3-flavor neutrino mixing framework that includes matter effects~\cite{ref:3flav} and introduces a dependency on the neutrino mass hierarchy, the sign of \mdmatm. The contours are calculated using the MINOS best fit values of $|$\dmsq{32}$|$ =$2.43\times10^{-3} {\rm eV^{2}}$ and \sinsq{23}=1.0. Statistical (poisson) and systematic effects (gaussian) are incorporated via the Feldman-Cousins approach which we use to determine the desired confidence intervals~\cite{ref:FC}. 

In conclusion we have described the first results of a search for \nue{} appearance in the MINOS experiment. The 1.5$\sigma$ excess of events can be interpreted in terms of \numunue{} oscillations. At 90\% CL we obtain an upper limit range of  $\sin^2(2\theta_{13})< 0.28-0.34$ for the normal neutrino mass hierarchy and $\sin^2(2\theta_{13})< 0.36-0.42$ for the inverted hierarchy depending upon $\delta_{CP}$.

MINOS is actively taking data and has already doubled the statistics used in this analysis (Spring 2009).  In the near future expect an improved analysis taking advantage of the larger statistics ($>7\times 10^{20}$~POT ) and improvements in our understanding of the systematic errors. 

\section*{Acknowledgments}
This work was supported by the US DOE; the UK STFC; the US NSF; the State and University of Minnesota; the University of Athens, Greece; and Brazil's FAPESP and CNPq.  We are grateful to the Minnesota Department of Natural Resources, the crew of the Soudan Underground Laboratory, and the staff of Fermilab for their contribution to this effort.

\end{document}